\documentclass[sigconf,natbib]{acmart}
\usepackage[nolist,nohyperlinks]{acronym}
\usepackage{multirow}
\AtBeginDocument{%
  \providecommand\BibTeX{{%
    \normalfont B\kern-0.5em{\scshape i\kern-0.25em b}\kern-0.8em\TeX}}}

\copyrightyear{2023}
\acmYear{2023}
\setcopyright{acmlicensed}\acmConference[SIGIR '23]{Proceedings of the 46th International ACM SIGIR Conference on Research and Development in Information Retrieval}{July 23--27, 2023}{Taipei, Taiwan}
\acmBooktitle{Proceedings of the 46th International ACM SIGIR Conference on Research and Development in Information Retrieval (SIGIR '23), July 23--27, 2023, Taipei, Taiwan}
\acmPrice{15.00}
\acmDOI{10.1145/3539618.3591951}
\acmISBN{978-1-4503-9408-6/23/07}
\newcommand{\up}[1]{\textsuperscript{#1}}

\begin{document}




\title{Attention Mixtures for Time-Aware Sequential Recommendation}

\author{Viet-Anh Tran, Guillaume Salha-Galvan, Bruno Sguerra, Romain Hennequin}
\affiliation{
  \institution{Deezer Research, Paris, France}
  \city{}
  \country{}
}
\email{research@deezer.com}

\renewcommand{\shortauthors}{V.A. Tran, et al.}

\begin{acronym}
    \acro{DL}{deep learning}
    \acro{NLP}{natural language processing}
    \acro{MF}{Matrix Factorization}
    \acro{HR}{Hit Rate}
    \acro{DCG}{Discounted Cumulative Gain}
    \acro{NDCG}{Normalized Discounted Cumulative Gain}
    \acro{CF}{Collaborative Filtering}
    \acro{FISM}{Factored Item Similarity Model}
    \acro{RNN}{recurrent neural networks}
    \acro{GRU}{Gated Recurrent Unit}
    \acro{LSTM}{Long Short-Term Memory}
    \acro{CNN}{convolutional neural networks}
    \acro{GNN}{graph neural networks}
    \acro{SR}{Sequential recommendation}
    \acro{MHA}{Multi-head attention}
    \acro{MOJITO}{Mixture Of Joint Item and Temporal cOntext}
\end{acronym}

\begin{abstract}
Transformers emerged as powerful methods for sequential recommendation. However, existing architectures often overlook the complex dependencies between user preferences and the temporal context. In this short paper, we introduce MOJITO, an improved Transformer sequential recommender system that addresses this limitation. MOJITO leverages Gaussian mixtures of attention-based temporal context and item embedding representations for sequential modeling. Such an approach permits to accurately predict which items should be recommended next to users depending on past actions and the temporal context. We demonstrate the relevance of our approach, by empirically outperforming existing Transformers for sequential recommendation on several real-world datasets. 
\end{abstract}





\maketitle

\section{Introduction}
\label{intro}
 \ac{SR} is essential to online platforms such as movie and music streaming services~\cite{Chen2018RecSysContinuation,schedl2018current,kang_icdm18,bontempelli2022flow,quadrana_csur21,sun_cikm19}. 
By modeling temporal sequences of user actions~\cite{fang_tois20, wang_csur21,quadrana_csur21}, SR systems predict the best content to recommend at a given time based on past interactions, e.g., the best songs to extend a listening session~\cite{pereira2019online}.
Over the past years, researchers have devoted significant efforts to improving \ac{SR}  \cite{hidasi_iclr16, trinh_recsys17, hu_sigir20, zhou_aaai18, you_www19, li_wsdm20, fang_tois20,ren_aaai19}. In particular, Transformers~\cite{vaswani_nips17}, the most recent class of neural architectures for sequence modeling, emerged as powerful \ac{SR} methods~\cite{kang_icdm18, sun_cikm19, zhang_aaai19, wu_recsys20, xie_arxiv20, jing_recsys20,du_cikm2022}.

However, most existing Transformer systems for SR omit or insufficiently exploit the \textit{temporal context} associated with user actions \cite{hansen_recsys20, li_wsdm20, cho_recsys20,cho_www21,ye_sigir20}. While they process the relative position of an interaction within a sequence, these systems often overlook time-related information such as the hour or the day of each interaction. Yet, such factors can be crucial in real-world \ac{SR} problems. For instance, in the music domain, Hansen et al. \cite{hansen_recsys20} observed that capturing temporal elements such as the hour of the day helps predict the following songs users listen to on a music streaming service. In another study focusing on various domains, from music to e-commerce, Ye et al.~\cite{ye_sigir20} identified several time-dependent patterns in interaction sequences. This includes seasonal behaviors, e.g., Christmas songs being more consumed in December, and periodic behaviors, e.g., buying a new toothbrush every three months.

This paper introduces \ac{MOJITO}, an improved Transformer SR system overcoming this limitation.
\ac{MOJITO} relies on the assumption that users tend to consume similar items, content-wise, for a given temporal context. Specifically, it jointly considers user-item interactions and the temporal context to learn two different attention matrices subsequently combined within Gaussian mixtures of temporal context and item embedding representations. This allows for an effective time-aware sequential modeling of user actions. We provide comprehensive
experiments on various real-world datasets covering application domains ranging from movie to music recommendation. 
Our results emphasize that \ac{MOJITO} empirically outperforms existing Transformers
for \ac{SR}, including models previously proposed to capture the dependencies between user preferences and the temporal context~\cite{li_wsdm20, cho_recsys20, cho_www21, rashed_recsys22}. Consequently, we believe \ac{MOJITO} will be helpful for researchers and practitioners eager to thoroughly leverage time-related contextual aspects when tackling \ac{SR} problems. We publicly release our code on GitHub to facilitate the~usage~of~our~system.

This paper is organized as follows. In Section~\ref{section2}, we formally present the \ac{SR} problem and review related work.
In Section~\ref{section3}, we introduce our proposed system for time-aware \ac{SR}.
We discuss our experimental analysis in Section~\ref{section4}, and conclude in Section~\ref{section5}.

\section{Preliminaries}
\label{section2}

\subsection{Problem Formulation}
\label{problem_formulation}

We consider a set~$\mathcal{U}$ of users and a set~$\mathcal{V}$ of items on an online service. For each $u~\in~\mathcal{U}$, we observe $n_u \geq 1$ \textit{sequences of interactions} on the service, e.g., listening sessions on a music streaming service. 
We denote by
$S_{ui} = (v_1, v_2, \dots, v_{L})$
the $i$\up{th} interaction sequence of~$u$, with $i \in \{1, \dots, n_u\}$, $v_t \in \mathcal{V}, \forall t \in \{1,\dots,L\}$, and a length\footnote{
Like Kang and McAuley~\cite{kang_icdm18}, we fix the length $L$ to simplify our presentation. Our work easily extends to sequences of varying lengths $L_{ui} < L$, by considering that the $L-L_{ui}$ first elements of such sequences are~null~values, acting as ``padding'' items~\cite{kang_icdm18}.} $L \in \mathbb{N}^*$.

Along with each $S_{ui}$, we observe a \textit{contextual sequence} $C_{ui}=(\mathbf{c}_1, \mathbf{c}_2, \dots, \mathbf{c}_{L})$. Each element $\mathbf{c}_{t}$ (with $t \in \{1,\dots, L\}$) is a tuple associating the $t$\up{th} interaction of $S_{ui}$  with $C \in\mathbb{N}^*$ different types of \textit{time-related contextual information}, such as the month (from January to December), the day of the month (from 1 to 31), or the day of the week (from Monday to Sunday) of this interaction. Formally, we set $\mathbf{c}_{t} = (c_{t1}, c_{t2},\dots, c_{tC})$, where $c_{tj} \in \mathcal{C}_j, \forall j\in\{1,\dots,C\}$, with $\mathcal{C}_j$ denoting the set of possible values for the $j$\up{th} context type. We order these values in each set $\mathcal{C}_j$ to account for temporal proximity, e.g., to capture that January is closer to February than to July. 

In such a setting, SR consists in predicting the next item $v_{L+1}$ a user will interact with, in the observed temporal context $\mathbf{c}_{L+1}$.

\subsection{Related Work}
\label{relatedwork}


Addressing an \ac{SR} problem usually requires to~jointly~account~for:
\begin{enumerate}
    \item ``Long-term'' preferences: this term designates intrinsic user preferences, independent of the current~context~\cite{wang_csur21}.
    \item ``Short-term'' preferences: on the contrary, this term refers to how the context affects preferences, including how recent interactions alter the perception of a recommendation~\cite{fang_tois20,adomavicius2011context}.
\end{enumerate}
In recent years, several \ac{SR} systems have been proposed to model these two aspects, often building upon advances in deep learning for sequence modeling~\cite{hidasi_iclr16, trinh_recsys17, hu_sigir20, zhou_aaai18, zhang_aaai19, you_www19, kang_icdm18, li_wsdm20, fang_tois20,ren_aaai19,guo2022reinforcement,zhang2022enhancing}.
This paper focuses on Transformers~\cite{vaswani_nips17} for SR. Using attention mechanisms~\cite{vaswani_nips17} for sequence modeling, they have recently shown competitive results on SR problems~\cite{kang_icdm18, sun_cikm19, zhang_aaai19, wu_recsys20, xie_arxiv20, jing_recsys20, du_cikm2022, yang2022multi, xia2022multi, chen2021learning}. 

\subsubsection{Transformers for SR}
SASRec~\cite{kang_icdm18} was the first system to leverage self-attention to identify relevant
items among temporal sequences for \ac{SR}. BERT4Rec~\cite{sun_cikm19} subsequently implemented bidirectional self-attention techniques.
To combine long-term and short-term preferences more carefully, AttRec~ \cite{zhang_aaai19} adopted a multi-task metric learning framework, while SSE-PT~\cite{wu_recsys20} concatenated  user ``embedding'' vector representations with each item in the sequence. Lately, FISSA~\cite{jing_recsys20} incorporated an attentive version of the \ac{FISM} \cite{kabbur_kdd13} for long-term user preference, and used an item similarity-based gating mechanism to merge short-term and long-term representations, with promising performances. Other recent work also combined attention mechanisms for \ac{SR} with graph neural networks~\cite{qiu_cikm19,wu_aaai19} or models based on contrastive learning~\cite{du_cikm2022,new3}, studied fairness-aware attention-based \ac{SR}~\cite{li2022fairsr}, and presented more effective training~strategies~\cite{petrov2022effective}. 



\subsubsection{Transformers for Time-Aware SR} 
Despite their success, these Transformers were also criticized for not fully exploiting \textit{temporal} factors \cite{hansen_recsys20, li_wsdm20, cho_recsys20,cho_www21,ye_sigir20}. While their attention mechanisms exploit relative positions in interaction sequences for preference modeling,
they neglect the temporal context of each interaction, captured by $C_{ui}$ sequences in Section~\ref{problem_formulation}. 
Yet, as illustrated in the introduction, time-related factors such as the current month, day, or hour can significantly modify user needs and preferences~\cite{bontempelli2022flow,hansen_recsys20,ye_sigir20}. 

Attempting to tackle this problem, TiSASRec~\cite{li_wsdm20} enhanced SASRec by analyzing time intervals between interactions using self-attention operations. TASER~\cite{ye_sigir20} combined pairwise relative and pointwise absolute time patterns in the same model. MEANTIME \cite{cho_recsys20} leveraged multiple self-attention heads to learn several embedding representations, encoding different temporal patterns. TimelyRec~\cite{cho_www21} modeled heterogeneous temporal patterns in user behaviors, by processing hierarchical time granularities. Recently, CARCA~\cite{rashed_recsys22} relied on contextual correlations computed via cross-attention to provide time-aware recommendations.

%
Nonetheless, we believe that time-aware SR remains incompletely addressed. While existing time-aware Transformers gather item and temporal representations using simple concatenations or additions, such an approach oversimplifies the complex relationship between user preferences and the temporal context.
Moreover, except MEANTIME~\cite{cho_recsys20}, the aforementioned methods suffer from redundant attention heads problems \cite{michel_neurips19, voita_acl19, bhojanapalli_arxiv21}. The remainder of this paper will emphasize that overcoming these limitations permits to better model preferences and empirically improve~time-aware~SR.

\section{A Mixture System for Time-Aware SR}
\label{section3}


\subsection{General Overview} 
\label{overview}
\begin{figure*}[t]
  \centering
  \includegraphics[width=1.0\textwidth]{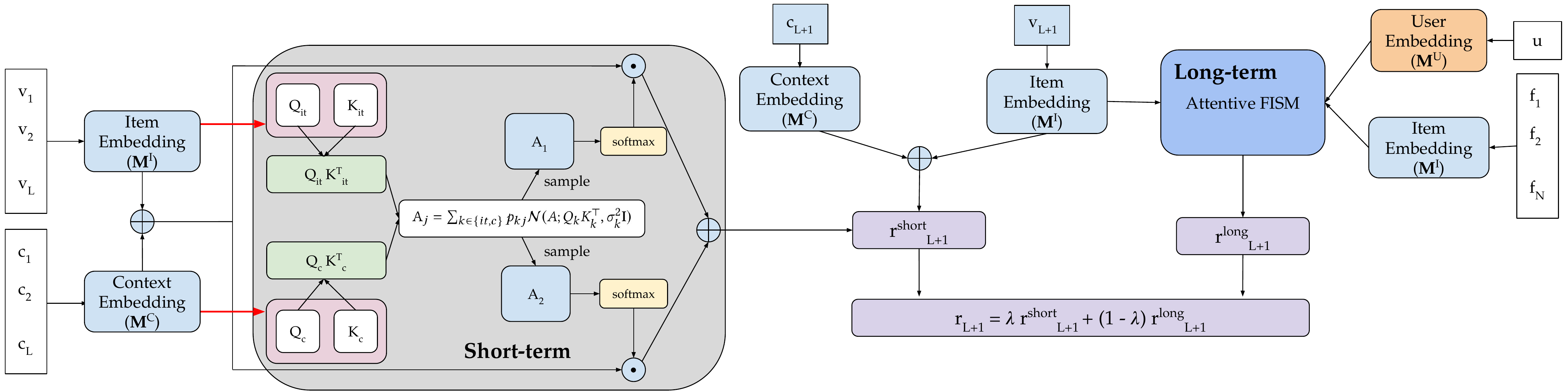} 
  \caption{Architecture of MOJITO for time-aware SR using attention mixtures of
temporal context and item embeddings.}
  \label{fig:mojito}
\end{figure*}

This Section~\ref{section3} introduces \ac{MOJITO}. As illustrated in Figure~\ref{fig:mojito}, our system involves two components. 
The first one, described in Section~\ref{short}, models short-term intents influenced by interactions and the temporal context. The second one, presented in Section~\ref{long}, captures long-term preferences. Each component returns a \textit{relevance score}, denoted $r^{\text{short}}_{L+1}(v)$ and $r^{\text{long}}_{L+1}(v)$ respectively. They estimate how likely each item $v\in\mathcal{V}$ would fit as an extension of an observed interaction sequence. To select the items to recommend, we combine scores as follows, for some hyperparameter $\lambda \in [0, 1]$:
\begin{equation}
 r_{L+1}(v) = \lambda r^{\text{short}}_{L+1}(v) + (1 - \lambda) r^{\text{long}}_{L+1}(v).
 \label{eq1}
\end{equation}
%

%


\subsection{Short-Term Representation}
\label{short}
We leverage self-attention mechanisms \cite{vaswani_nips17, kang_icdm18}
for time-aware short-term representation learning.
As this approach is known to suffer from redundant heads\footnote{We refer, for instance, to the experiments of Kang and McAuley~\cite{kang_icdm18}, showing that increasing the number of heads in SASRec does not improve sequential recommendation.} problems where different attention heads actually learn the same information \cite{michel_neurips19, voita_acl19,bhojanapalli_arxiv21}, 
we build upon recent efforts from Nguyen et al. \cite{nguyen_neurips22} and enable our attention heads to interact via a shared pool of global attention matrices. Nguyen et al. \cite{nguyen_neurips22} sampled the local attention matrix of each head using an admixture model \cite{blei_jmlr03}. In our work, we enhance the diversity in the representation of global matrices by introducing different inductive semantic biases to guide their learning, as~described~below.

\subsubsection{Embedding Layer}
This first block processes the $S_{ui}$ and $C_{ui}$ sequences from Section~\ref{problem_formulation}.
We denote by $\mathbf{M}^I \in \mathbb{R}^{|\mathcal{V}| \times d}$ a learnable \textit{item embedding matrix}, in which rows are $d$-dimensional embedding vectors representing each item, for some dimension $d \in \mathbb{N}^*$. Using such a notation, a sequence $S_{ui}$ can be represented as an embedding matrix, as follows: $\mathbf{E}^I_{S_{ui}} = [ \mathbf{m}^I_{v_1}, \mathbf{m}^I_{v_2},\dots, \mathbf{m}^I_{v_L} ]^{\intercal} \in \mathbb{R}^{L \times d}$. 

We leverage the \textit{translation-invariant time kernel} of Xu et al. \cite{xu_neurips19} to learn $d$-dimensional \textit{temporal context embedding vectors} compatible with self-attention operations. 
Specifically, each $c_{ti} \in \mathbf{c}_t$ is mapped to a vector $\mathbf{m}^C_{c_{ti}} \in \mathbb{R}^d$ using a Mercer kernel function \cite{xu_ijcai19}. We concatenate and inject these vectors into a linear layer function $g(\cdot)$ to obtain an embedding representation $\mathbf{m}^C_{\mathbf{c}_t} = g([\mathbf{m}^C_{c_{t1}}; \mathbf{m}^C_{c_{t2}};\dots;\mathbf{m}^C_{c_{tC}}]) \in \mathbb{R}^d$ for each context tuple $\mathbf{c}_t$. We obtain an \textit{item/context embedding matrix} combining item and context embeddings: $
\mathbf{E}_{S_{ui},C_{ui}} = 
[[\mathbf{m}^I_{v_1}; \mathbf{m}^C_{\mathbf{c}_1}];\dots; [\mathbf{m}^I_{v_L}; \mathbf{m}^C_{\mathbf{c}_L}]]^{\intercal} \in \mathbb{R}^{L \times 2d} $. 

Lastly, to model the influence of positions in sequences, we enrich $\mathbf{E}_{S_{ui},C_{ui}}$ with learnable \textit{position embeddings} $\mathbf{P} = [ \mathbf{p}_1, \dots, \mathbf{p}_L ]^{\intercal} \in \mathbb{R}^{L \times 2d}$, obtaining our final \textit{input matrix} $\mathbf{X}^{(0)} = [ \mathbf{x}^{(0)}_1, \dots, \mathbf{x}^{(0)}_L]^{\intercal} \in \mathbb{R}^{L \times 2d}$, where $\mathbf{x}^{(0)}_l = [\mathbf{m}^I_{v_l}; \mathbf{m}^C_{\mathbf{c}_l}] + \mathbf{p}_l, \forall l \in \{1,\dots, L\}$. We aim to distinguish different influences of the same item in different~positions.

\subsubsection{Attention Mixtures}
We pass $\mathbf{X}^{(0)}$ through $B \in \mathbb{N}^*$ stacked \textit{self-attention blocks} (SAB)~\cite{vaswani_nips17}. The output of the $b$\up{th} block is $\mathbf{X}^{(b)} = \text{SAB}^{(b)}(\mathbf{X}^{(b-1)})$, for $b \in \{1, \dots, B\}$. 
The SAB contains a \textit{self-attention layer} $\text{SAL}(\cdot)$ with $H \in \mathbb{N}^*$ heads mixing context and item embedding representations, followed by a \textit{feed-forward~layer}~$\text{FFL}(\cdot)$:

\noindent
\begin{align}
    \text{SAL}(\mathbf{X}) &= \text{MultiHead}(\{\mathbf{X}^{\text{Att}}_j\}^H_{j=1}) = \text{Concat}(\mathbf{X}^{\text{Att}}_1,\dots,\mathbf{X}^{\text{Att}}_H) \mathbf{W}^O, \nonumber \\
    \text{SAB}(\mathbf{X}) &= \text{FFL}(\text{SAL}(X)) = ReLU(\mathbf{X}^{\text{Att}} \mathbf{W}_1 + \mathbf{b}_1) \mathbf{W}_2 + \mathbf{b}_2,
\end{align}
where $\mathbf{X} \in \mathbb{R}^{L \times 2d}$ is the input of each block, $\mathbf{W}^O \in \mathbb{R}^{2Hd \times 2d}$ is the projection matrix for the output, and $\mathbf{W}_1$, $\mathbf{W}_2 \in \mathbb{R}^{2d \times 2d}$ and $\mathbf{b}_1, \mathbf{b}_2 \in \mathbb{R}^{1 \times 2d}$ are weights and biases for the two layers of the FFL network. $\mathbf{X}^{\text{Att}}_j = \text{softmax}(\mathbf{A_j} / \sqrt{d}) \mathbf{V}$ is the output of each head. In this work, we sample $\mathbf{A_j}$ from a \textit{Gaussian mixture probabilistic model}, automatically learning the relative importance of the temporal context and previous item interactions for sequential modeling:
\begin{align}
    \mathbf{A}_j &\sim \sum_{k \in \{it, c\}} p_{kj} \mathcal{N}(\mathbf{A}; \mathbf{Q}_k \mathbf{K}^\intercal_k, \sigma^2_k), \sum_{k \in \{it, c\}} p_{kj} = 1, p_{kj} \geq 0,
\end{align}
where $\mathbf{Q}_k = \mathbf{X}_k\mathbf{W}^k_Q$, $\mathbf{K}_k = \mathbf{X}_k\mathbf{W}^k_K$ and $\mathbf{V} = \mathbf{X}\mathbf{W}_V, $ with $\mathbf{W}^k_Q, \mathbf{W}^k_K \in \mathbb{R}^{d \times d}$ and $\mathbf{W}_V \in \mathbb{R}^{2d \times 2d}$ are the projected value matrices. The variance term $\sigma^2_k$, learned by the model, is applied element-wise to each element of the corresponding attention matrix.



\subsubsection{Prediction}
\label{st-pred}
We use a \textit{latent factor model}~\cite{kang_icdm18,li_wsdm20} to estimate the relevance of any $v \in \mathcal{V}$ to extend $S_{ui}$, in the observed context $\mathbf{c}_{L+1}$. We obtain the \textit{short-term relevance score} as follows:
\begin{equation}
r^{\text{short}}_{L+1}(v) =  \mathbf{x}^{(B)\intercal}_L~[\mathbf{m}^{I}_{v}; \mathbf{m}^{C}_{\mathbf{c}_{L+1}}],
\end{equation}
where $\mathbf{x}^{(B)}_{L}$ is the model output at position $L$ after $B$ attention blocks. 

\subsection{Long-Term Representation}
\label{long}

We now turn to long-term representation learning. At this stage, we found it crucial to capture the \textit{heterogeneity} of user preferences. 

\subsubsection{Attentive FISM} In real-world applications, users often have diverse preferences \cite{tay_www18, park_icdm18, zhang_dsaa19}. For instance, one can simultaneously like metal and classical music.
Hence, the practice of representing long-term user tastes using a single embedding vector (as in AttRec \cite{zhang_aaai19} or SSE-PT \cite{wu_recsys20}) has been shown to be limiting when preferences are dispersed in the embedding space~\cite{tay_www18,tran_recsys21}. 

 To overcome this issue, we use an attentive version of \ac{FISM}~\cite{kabbur_kdd13} to represent users in MOJITO. This model, also incorporated in the FISSA system for SR \cite{jing_recsys20}, allows user representations to flexibly \textit{translate} depending on the current target item representation~\cite{tran_recsys21}. 
Contrary to FISSA, we learn user representations using $N \in \mathbb{N}^*$ randomly selected items from the user's past history, instead of using only items from the currently observed sequence. We aim to model more diverse long-term interests by adopting such a strategy. 

Formally, denoting by $\mathcal{F} = \{f_1,\dots,f_N\}$ the $N$ selected items, the long-term preference of user $u$ for an item $v \in \mathcal{V}$ at any time $l+1$ is estimated using a weighted aggregation of item representations: 

\noindent
\begin{equation}
    \mathbf{\tilde{m}}_u (v)  = \mathbf{m}_u + \sum_{f \in \mathcal{F}~ \setminus \{v\}} \frac{e^{\mathbf{m}_{f}^\intercal \mathbf{m}_{v}}}{\sum_{f' \in \mathcal{F}~\setminus \{v\}} e^{\mathbf{m}_{f'}^\intercal \mathbf{m}_{v}}} \mathbf{m}_{f} \in \mathbb{R}^d,
\end{equation}
where the vector $\mathbf{m}_u \in \mathbb{R}^d$ is learned by the model during training.

\subsubsection{Prediction}
Similarly to Section~\ref{st-pred}, we obtain \textit{long-term relevance scores} as follows:
\begin{equation}
r^{\text{long}}_{L+1} (v) = \mathbf{m}^\intercal_{v}~\mathbf{\tilde{m}}_u (v).
\end{equation}
We emphasize that this operation does \textit{not} involve any temporal context information, as long-term preferences are assumed to be independent of the~context by definition (see Section~\ref{relatedwork}).

\subsection{Training Procedure}
\label{opti}

MOJITO involves various weight matrices to optimize. For this purpose, we consider a training set $\mathcal{S}$ of interaction and context sequences.
For each interaction sequence $S_{ui}$, we create sub-sequences composed of the $l$ first items of $S_{ui}$ for $l \in \{1,\dots, L\}$. We denote by $v_{S_{ui}, l+1}$ the $(l+1)$\up{th} ground truth item extending each sequence. Additionally, we sample a ``negative'' item $o_{S_{ui}, l+1} \in \mathcal{V} \setminus v_{S_{ui}, l+1}$ for each sub-sequence.
We expect our system to return high relevance scores
for $v_{S_{ui}, l+1}$ items, and lower relevance scores for negative items. To achieve this, we optimize model weights via gradient descent minimization of the loss $\mathcal{L} = \lambda \mathcal{L}^{\text{short}} + (1 - \lambda) \mathcal{L}^{\text{long}}$ (we recall that $\lambda \in [0, 1]$), a linear combination of the $\mathcal{L}^{\text{short}}$ and $\mathcal{L}^{\text{long}}$ losses defined as follows: %
\begin{equation}
    \label{loss}
    \small \mathcal{L}^{x} = - \sum_{S \in \mathcal{S}} \sum_{l=1}^{L} \left[\log\left(\sigma(r^{x}_{l+1}(v_{S, l+1}))\right) + \log\left(1 - \sigma(r^{x}_{l+1}(o_{S, l+1}))\right) \right],
\end{equation}
for $x \in \{\text{short}, \text{long}\}$ and the sigmoid function $\sigma(a) = \frac{1}{1+e^{-a}}$. 


\begin{table*}[h!]
  \caption{Sequential Recommendation (SR) on all datasets using MOJITO and other baselines.  Scores are computed on test items. Models recommend ranked lists of 10 items. Bold numbers are the best scores. Underlined numbers are the second-best~ones.}
  \label{all_results}
  \resizebox{1.0\textwidth}{!}{
  \begin{tabular}{c|c||cc|cc|cc}
        \toprule
        \multirow{2}{*}{\textbf{Type}} & \multirow{2}{*}{\textbf{Model}} & \multicolumn{2}{c}{\textbf{MovieLens} {\small ($k^\text{user} =$ 10, $k^\text{item} =$ 5)}} & \multicolumn{2}{c}{\textbf{Amazon Book} {\small ($k^\text{user} =$ 30, $k^\text{item} =$ 20)} } & \multicolumn{2}{c}{\textbf{LFM-1b} {\small ($k^\text{user} =$ 300, $k^\text{item} =$ 500)}} \\
        
        \cline{3-8}
        {} & {} & NDCG (in \%) & HR (in \%) & NDCG (in \%) & HR (in \%) & NDCG (in \%) & HR (in \%) \\
        \midrule
        \multirow{5}{*}{\shortstack{Non Time-Aware}} & {SASRec} & {58.27 $\pm$ 0.14} & {80.86 $\pm$ 0.46} & {55.23 $\pm$ 0.16} & {78.29 $\pm$ 0.19} & {\underline{\textit{58.01 $\pm$ 0.22}}} & {73.84 $\pm$ 0.19} \\
        {} & {BERT4Rec} & {53.75 $\pm$ 0.17} & {75.59 $\pm$ 0.18} & {57.17 $\pm$ 0.15} & {79.90 $\pm$ 0.08} & {56.08 $\pm$ 0.04} & {68.87 $\pm$ 0.06} \\
        {} & {SSE-PT} & {55.62 $\pm$ 0.24} & {79.61 $\pm$ 0.19} & {55.00 $\pm$ 0.25} & {79.63 $\pm$ 0.11} & {56.60 $\pm$ 0.39} & {\underline{\textit{75.21 $\pm$ 0.36}}} \\
        {} & {AttRec} & {42.08 $\pm$ 0.31} & {69.23 $\pm$ 0.21} & {44.67 $\pm$ 0.20} & {71.00 $\pm$ 0.24} & {43.81 $\pm$ 0.19} & {64.66 $\pm$ 0.38} \\
        {} & {FISSA} & {48.53 $\pm$ 0.31} & {74.16 $\pm$ 0.36} & {58.94 $\pm$ 0.11} & {81.42 $\pm$ 0.10} & {52.40 $\pm$ 0.19} & {68.20 $\pm$ 0.23} \\
        \hline
        \multirow{4}{*}{Time-Aware} & {TiSASRec} & {58.09 $\pm$ 0.26} & {\underline{\textit{80.86 $\pm$ 0.28}}} & {52.16 $\pm$ 0.07} & {78.53 $\pm$ 0.22} & {55.67 $\pm$ 0.33} & {73.25 $\pm$ 0.37} \\
        {} & {MEANTIME} & {\textbf{59.97 $\pm$ 0.18}} & {79.76 $\pm$ 0.15} & {\underline{\textit{58.95 $\pm$ 0.11}}} & {82.02 $\pm$ 0.11} & {57.03 $\pm$ 0.06} & {71.60 $\pm$ 0.07} \\
        {} & {CARCA} & {38.65 $\pm$ 0.14} & {64.51 $\pm$ 0.09} & {56.82 $\pm$ 0.23} & {\underline{\textit{82.89 $\pm$ 0.17}}} & {49.38 $\pm$ 0.44} & {66.30 $\pm$ 0.38} \\
        \cline{2-8}
        {} & {MOJITO (ours)} & {\underline{\textit{59.82 $\pm$ 0.17}}} & {\textbf{82.09 $\pm$ 0.22}} & {\textbf{59.94 $\pm$ 0.23}} & {\textbf{83.26 $\pm$ 0.09}} & {\textbf{60.14 $\pm$ 0.19}} & {\textbf{75.72 $\pm$ 0.23}} \\
        \bottomrule
      \end{tabular}
    }
\end{table*}

\section{Experimental Analysis}
\label{section4}
This Section~\ref{section4} presents our experimental evaluation of MOJITO. For reproducibility and future usage, we release our code on GitHub\footnote{\label{sourcecode}\href{https://github.com/deezer/sigir23-mojito}{https://github.com/deezer/sigir23-mojito}}.

\subsection{Experimental Setting}

\subsubsection{Datasets}
We consider three real-world datasets covering different domains, namely movie, book, and music recommendation:

\begin{enumerate}
    \item MovieLens \cite{harper2015movielens}: a movie recommendation dataset gathering 1 million ratings of 3~883 movies by 6~040 users. Each rating acts as a binary ``interaction'' in the following.
    \item Amazon Book \cite{mcauley_sigir15}: a collection of more than 1 million interactions with  96~421 books by 109~730 Amazon users. 
    \item LFM-1b \cite{schedl_icmr16}: more than a billion listening events by 120~322 Last.fm users, associated with 3~190~371 distinct music items.
\end{enumerate}
Like Sun et al.~\cite{sun_recsys20}, we implement a $k^{\text{item}}$-core (respectively, a $k^{\text{user}}$-core) pre-processing step on each dataset, 
consisting in recursively filtering data until all users (resp., all items) have at least $k^{\text{item}}$ (resp., $k^{\text{user}}$) interactions. Values of $k^{\text{item}}$ and $k^{\text{user}}$ are reported in Table~\ref{all_results}. 


\subsubsection{Task} For each interaction sequence, the last item is held out for test and the penultimate one for validation. We evaluate the ability of SR models to retrieve these items in evaluation sets containing the ground truth missing item and 1K other ``negative'' items that the user has not interacted with. SR models must recommend a list of 10 items, ordered by relevance scores. We evaluate recommendations using the Hit Rate (HR@10) and Normalized Discounted Cumulative Gain (NDCG@10) metrics. The former metric reports the percentage of top-10 lists containing missing items, whereas the latter also takes into account the missing item's ranking in~the~list.

\subsubsection{Baselines}
We compare MOJITO to eight of the best Transformer models mentioned in Section~\ref{relatedwork}: 
SASRec \cite{kang_icdm18}, BERT4Rec \cite{sun_cikm19},
SSE-PT \cite{wu_recsys20}, FISSA \cite{jing_recsys20}, AttRec \cite{zhang_aaai19}, TiSASRec \cite{li_wsdm20}, MeanTime \cite{cho_recsys20}, and CARCA \cite{rashed_recsys22}. We recall that, like MOJITO, the three last baselines explicitly aim to capture temporal aspects for SR.

\subsubsection{Implementation Details}

We train all models for a maximum of 100 epochs using the Adam optimizer \cite{kingma_iclr15}. 
We set $d =$ 64, $L =$~50, batch sizes of 512, and $B =$ 2, $H =$ 2 for MOJITO. 
We selected all other hyperparameters via a grid search on validation items. For brevity, we report optimal values of all models in our GitHub repository\textsuperscript{\ref{sourcecode}}. Most notably, we tested learning rates values in \{0.0002, 0.0005, 0.00075, 0.001\}, $N$ in \{20, 50, 100\}, and $\lambda$ in \{0.1, 0.2, 0.5,~0.8,~1.0\}.

\subsection{Results and Discussion}
Table \ref{all_results} reports all results on test sets, along with standard deviations computed over five model trainings. Overall, MOJITO reaches competitive results on all three datasets, which we~discuss~below.

\subsubsection{Time-Aware vs Non Time-Aware SR}
\label{s421}
Our analysis reveals that time-aware systems tend to outperform non time-aware baselines, especially on MovieLens and Amazon Book. This highlights the relevance of modeling dependencies between preferences and the temporal context for SR in various domains. Yet, these performance differences are less pronounced on LFM-1b, likely due to the complexities of the temporal context in music consumption. Music consumption has lower levels of engagement than book or movie consumption, resulting in increased frequency of interactions and shorter time intervals between them \cite{schedl2019deep}. This makes modeling the temporal context for music recommendation more challenging, especially given the higher number of music tracks for the same context. Nonetheless, \ac{MOJITO}, explicitly accounting for day-of-the-week periodic information, has the best performance on LFM-1b (60.14\% NDCG, 75.72\%~HR), showing its capacity to model the granularity of music consumption that is prone to~periodic~patterns~\cite{bontempelli2022flow}.

\subsubsection{MOJITO vs Other Time-Aware Systems}
While accounting for the temporal context is crucial, the approach used to achieve this goal is equally important.
In our experiments,~\ac{MOJITO} outperforms time-aware baselines according to 5 out of 6 scores (e.g., with a top 83.26\% HR and a top 59.94\% NDCG on Amazon Book) and reaches the second-best result for the 6\up{th} score (i.e., 59.82\% NDCG on MovieLens, close to the top 59.97\% NDCG obtained with MEANTIME on this dataset). 
These positive results emphasize the effectiveness of our flexible mixture-based probabilistic approach to automatically learn complex relationships between the temporal context and user preferences from various application domains.

\subsubsection{Future Analyses}
To conclude our discussion, we formulate two additional conjectures, that we plan to investigate in future research.
Firstly, we believe that the inferior performance of CARCA with respect to time-aware alternatives on MovieLens and LFM-1B in Table~\ref{all_results} could also be explained by the relative complexity of this system. Indeed, CARCA uses an entire Transformer encoder-decoder architecture while others only employ the decoder part~\cite{rashed_recsys22}.
Secondly, we recall that, by design, \ac{MOJITO} aims to reduce head redundancies, which our experiments confirm. For instance, we report a 0.85 $\pm$ 0.43 mean pairwise $L_2$ distance between head distances at the output layer for MOJITO on Amazon Book, vs. 0.57~$\pm$~0.13 for TiSASRec and 0.76 $\pm$ 0.59 for CARCA (large values indicate less head redundancy). This head diversity seems to contribute to performance gains, as in the work of Nguyen et al.~\cite{nguyen_neurips22}, although more ablation studies will be required in future work for confirmation.

\section{Conclusion}
\label{section5}



In this paper, we proposed an effective Transformer for time-aware SR using mixtures of attention-based temporal context and item embedding representations. Backed by conclusive experiments on real-world datasets from various domains, we demonstrated its ability to thoroughly capture the impact of the temporal context on user preferences, enabling a more careful sequential modeling of user actions for SR. 
We also identified areas of analysis that, in the future, would help us better understand and improve our~system.

\bibliographystyle{ACM-Reference-Format}
\bibliography{sample-base}




\end{document}